\documentclass[pdflatex,sn-mathphys-num]{sn-jnl}

\usepackage{graphicx}%
\usepackage{multirow}%
\usepackage{amsmath,amssymb,amsfonts}%
\usepackage{amsthm}%
\usepackage{mathrsfs}%
\usepackage[title]{appendix}%
\usepackage{xcolor}%
\usepackage{textcomp}%
\usepackage{manyfoot}%
\usepackage{booktabs}%
\usepackage{algorithm}%
\usepackage{algorithmicx}%
\usepackage{algpseudocode}%
\usepackage{listings}%
\usepackage{bm}%
\usepackage{subfigure}

\begin{document}

\title[Review of heat and charge transport in strongly magnetized relativistic plasmas]{Review of heat and charge transport in strongly magnetized relativistic plasmas}

\author*[1,2]{\fnm{Igor A.} \sur{Shovkovy}
}\email{igor.shovkovy@asu.edu}
\equalcont{These authors contributed equally to this work.}

\author[1]{\fnm{Ritesh} \sur{Ghosh}
}\email{ritesh.ghosh@asu.edu}
\equalcont{These authors contributed equally to this work.}

\affil*[1]{\orgdiv{College of Integrative Sciences and Arts}, \orgname{Arizona State University}, \orgaddress{\street{6073 S. Backus Mall}, \city{Mesa}, \postcode{85212}, \state{AZ}, \country{USA}}}

\affil*[2]{\orgdiv{Department of Physics}, \orgname{Arizona State University}, \orgaddress{\street{550 E. Tyler Mall}, \city{Tempe}, \postcode{85287}, \state{AZ}, \country{USA}}}

\abstract{We review field-theoretic studies of charge transport in hot relativistic plasmas under strong magnetic fields and extend the analysis to thermal conductivity. The calculations rely on accurately determining the fermion damping rate. Using the Landau-level representation, these damping rates are computed exactly at leading order and incorporated into the Kubo formula to obtain the thermal and electrical conductivity tensors. Our analysis reveals that the mechanisms underlying longitudinal and transverse transport differ significantly. Strong magnetic fields markedly suppress transverse charge transport by confining particles within localized Landau orbits, allowing transport only through quantum transitions between these discrete states. In contrast, longitudinal charge transport is enhanced, as it primarily depends on the reduced scattering probability of particles moving along the direction of the magnetic field. The anisotropy of thermal conductivity is also nontrivial but less pronounced since its underlying transport mechanism is different. We also examine the modification of the Wiedemann--Franz law in strongly magnetized plasmas.}


\maketitle

\section{Introduction}\label{sec1}
 
Relativistic plasmas naturally arise in both astrophysical and cosmological settings. In laboratory environments, small droplets of hot quark-gluon plasma (QGP) are created in relativistic heavy-ion collisions \cite{STAR:2005gfr,PHENIX:2004vcz,PHOBOS:2004zne}, mimicking the conditions that have existed in the early Universe shortly after the Big Bang \cite{Yagi:2005yb}. 

Astrophysical plasmas are frequently permeated by strong magnetic fields. One extreme example is a magnetar \cite{Sturrock:1971zc,Ruderman:1975ju,Arons:1983aa,Turolla:2015mwa,Kaspi:2017fwg}, with surface magnetic fields reaching up to $10^{15}~\mbox{G}$ and potentially even stronger fields in the interior \cite{Lai_1991}. Such intense fields play a crucial role in stellar dynamics and evolution. Strong magnetic fields were also likely present in the early Universe \cite{Grasso:2000wj}, offering a natural explanation for the present day intergalactic magnetic fields \cite{Durrer:2013pga,Vachaspati:2020blt}. In heavy-ion collisions, very strong magnetic fields are generated transiently by the electric currents produced by the motion of colliding charged ions \cite{Skokov:2009qp,Voronyuk:2011jd,Deng:2012pc,Bloczynski:2012en,Guo:2019mgh}. The electrical conductivity of the resulting quark-gluon plasma governs how these magnetic fields decay or diffuse, and recent results from the STAR experiment~\cite{STAR:2023jdd} provide valuable constraints on the electrical conductivity of the QGP, encouraging further exploration of magnetic effects.

In this review, we summarize our recent first-principles investigations of the electrical conductivity of magnetized hot relativistic plasmas at the vanishing chemical potential \cite{Ghosh:2024fkg,Ghosh:2024owm} and extend the analysis to the thermal conductivity. Understanding such transport is vital not only theoretically but also experimentally and observationally. In astrophysics, accurate knowledge of thermal and electrical conductivities is essential for predicting the evolution of magnetic fields in neutron stars and magnetars, which is crucial for understanding their behavior, emissions, and impacts on surrounding space. For instance, electrical conductivity governs deviations from the ``force-free" condition often assumed in pulsar magnetospheres, thereby significantly affecting observational predictions. In heavy-ion collisions, understanding transport is crucial for interpreting experimental data and identifying phenomena related to strong magnetic fields and extreme relativistic effects. Such insights may also shed light on conditions in the early Universe, where magnetized relativistic plasmas briefly existed.

Although the thermal and electrical conductivities are fundamental transport properties, their behavior in the presence of strong magnetic fields remains only partially understood. In the past, many studies primarily focused on weak-field regimes \cite{vanErkelens:1984,Pike:2016aa}. Subsequent work on QGP in strong fields employed various approaches, including analytical methods \cite{Hattori:2016cnt,Hattori:2016lqx,Fukushima:2017lvb,Fukushima:2019ugr}, lattice simulations \cite{Buividovich:2010qe,Buividovich:2010tn,Astrakhantsev:2019zkr,Almirante:2024lqn}, holography \cite{Mamo:2012kqw,Fukushima:2021got}, and phenomenological models \cite{Nam:2012sg,Kerbikov:2014ofa,Satapathy:2021cjp,Satapathy:2021wex,Bandyopadhyay:2023lvk}. Kinetic theory has also been widely applied \cite{Kurian:2017yxj,Das:2019ppb,Thakur:2019bnf,Dey:2020awu,K:2022pzc,Shaikh:2022sky,Singh:2023ues}, though its effectiveness is limited in the strong-field regime, particularly due to the use of the relaxation-time approximation.

In the crusts of magnetized neutron stars, electron heat and charge transport has been extensively studied, e.g., see Refs.~\cite{Hernquist_1984,Schaaf:1988AA,Potekhin:1999ur,Harutyunyan:2016rxm,Harutyunyan:2023ooz}. In such environments, dissipation of electron transport is primarily due to scattering from ions and phonons. In contrast, this mechanism does not apply to quark-gluon or electron-positron plasmas, which lack ions and crystalline lattices. Instead, in such hot relativistic plasmas, dissipation is governed by quantum transitions between Landau levels, involving processes such as $\psi_{n} \to \psi_{n'} + \gamma$, $\psi_{n} + \gamma \to \psi_{n'}$, and $\psi_{n} + \bar{\psi}_{n'} \to \gamma$, where $\psi_n$ denotes a charged fermion in the $n$th Landau level and $\gamma$ represents a photon \cite{Ghosh:2024hbf}. In the case quark-gluon plasma, the photon is replaced by a gluon. 

Building on recent developments in quantum field theory in the presence of background magnetic fields, specifically, techniques employing the Landau-level representation of Green's functions \cite{Miransky:2015ava}, we computed the anisotropic electrical conductivity in strongly magnetized relativistic plasmas using Kubo's formalism \cite{Ghosh:2024fkg,Ghosh:2024owm}. The key input in this approach is the leading-order fermion damping rate, derived via first-principles methods and driven by quantum one-to-two and two-to-one processes \cite{Ghosh:2024hbf}. These rates determine the spectral properties of charge carriers and are used to evaluate the full conductivity tensor. The framework is directly applicable to weakly coupled gauge theories such as quantum electrodynamics (QED), and is expected to work for quantum chromodynamics (QCD) at sufficiently high temperatures, where the coupling becomes weak due to asymptotic freedom  \cite{Gross:1973id,Politzer:1973fx}.

As detailed in this review, the electrical conductivities of a relativistic plasma become highly anisotropic in the presence of a strong magnetic field \cite{Ghosh:2024fkg,Ghosh:2024owm}. Transverse charge transport (across magnetic field lines) is strongly suppressed, as the field confines charged particles to quantized Landau-level orbits. This suppression can only be lifted through interactions that induce transitions between these discrete states. In contrast, longitudinal charge transport (along magnetic field lines) is significantly enhanced by the field. Because particle motion in this direction is not restricted by the magnetic field, charged carriers can propagate almost freely. Furthermore, the longitudinal conductivity increases with the field strength, due to reduced scattering and enhanced coherence of charge flow along the field lines.
 
Here we extend the analysis of Refs.~\cite{Ghosh:2024fkg,Ghosh:2024owm} to describe heat transport, which also becomes anisotropic in the presence of a strong magnetic field, although to a much lesser degree than charge transport. It should be noted that the mechanism of heat conduction in relativistic plasmas at vanishing chemical potential, which is the focus of our study here, is rather unusual. Since the pressure in equilibrium is uniquely determined by the local temperature, one might expect that a purely thermal gradient cannot be established without simultaneously inducing mechanical motion. This does not mean that heat conduction is absent or meaningless. Rather, the apparent difficulty of its interpretation arises in large part from the conventional choice of the Landau-Lifshitz hydrodynamic frame, with its intrinsic limitations for viscous fluids \cite{Kovtun:2019hdm,Bemfica:2020zjp}. In other hydrodynamic frames, consistent with causality and stability requirements, thermal conductivity has observable manifestations even at vanishing chemical potential. From a physics viewpoint, the underlying mechanism is tied to the dissipative components of sound-like collective modes. Put simply, as such a mode passes, it heats and cools a local region, while its dissipation deposits a finite amount of energy (heat).

\section{Kubo's formalism for thermal and electrical conductivities}

Using Kubo's linear response theory, the electrical conductivity tensor $\sigma_{ij}$ can be expressed in terms of the retarded current-current correlation function~\cite{Ghosh:2024fkg,Ghosh:2024owm}: 
 \begin{equation}
  \sigma_{ij}  = \lim_{\Omega\to 0}\frac{\mbox{Im}\left[\Pi_{ij}^{\rm R}(\Omega+i0 ;\mathbf{0}) \right]}{\Omega}.
  \label{sigma-tensor}
 \end{equation}
For simplicity, we consider a QED-like plasma consisting of a single fermion species of mass $m$. The generalization to a multi-component system is straightforward. In the weak-coupling regime, using the one-loop expression for the current-current correlation function, we obtain the following result for the electrical conductivity tensor~\cite{Ghosh:2024fkg,Ghosh:2024owm}:
 \begin{equation}
  \sigma_{ij}  =  -\frac{\alpha}{8\pi T} \int  \frac{dk_0 d^3 \bm{k}}{\cosh^2\frac{k_0}{2T}}
  \mbox{tr} \left[ \gamma^i A_{\bm{k}} (k_0) \gamma^j A_{\bm{k}} (k_0) \right] ,
  \label{sigma-tensor-spectral}
 \end{equation}
 where $\alpha=e^2/(4\pi)$ is the (fine-structure) coupling constant and the fermion spectral function is defined as $A_{\bm{k}} (k_0)\equiv \left[\bar{G}(k_0-i0,\bm{k}) -\bar{G}(k_0+i0,\bm{k})\right]/(2\pi i)$, with $\bar{G}(k_0,\bm{k})$ denoting the Fourier transform of the translation-invariant part of the fermion propagator in the presence of a background magnetic field. In the Landau-level representation, the explicit expression of the spectral function reads \cite{Miransky:2015ava}
\begin{eqnarray}
A_{\bm{k}} (k_0) &=& i \sum_{\lambda=\pm}\sum_{n=0}^{\infty} 
\frac{(-1)^n}{E_n} e^{-k_\perp^2\ell^2}  
   \rho\left(k_0,\lambda E_n\right)
  \Big\{
  \left[E_{n} \gamma^{0} 
  -\lambda  k_{z}\gamma^3+\lambda  m \right]   \nonumber\\
  & \times & \left[{\cal P}_{+}L_n\left(2 k_\perp^2\ell^2\right)
  -{\cal P}_{-}L_{n-1}\left(2 k_\perp^2\ell^2\right)\right] + 2\lambda  (\bm{k}_\perp\cdot\bm{\gamma}_\perp) L_{n-1}^1\left(2 k_\perp^2 \ell^2\right)
  \Big\} ,   
  \label{spectral-density}
 \end{eqnarray}
where $E_{n}=\sqrt{2n|eB|+m^2+k_z^2}$ is the Landau-level energy, $\ell = 1/\sqrt{|eB|}$ is the magnetic length, ${\cal P}_{\pm}=\frac{1}{2}\left[ 1\pm i\, \mbox{sign}(e B) \gamma^1\gamma^2 \right] $
are spin projectors, $L_n^\alpha(z)$ are the generalized Laguerre polynomials~\cite{Gradshteyn:1980}, and the function $\rho\left(k_0,\lambda E_n\right)$ is a Lorentzian of width $\Gamma_n$, centered at $\lambda E_n$, i.e.,  
\begin{equation} 
\rho\left(k_0,\lambda E_n\right) = \frac{1}{\pi} \frac{\Gamma_n}{(k_0-\lambda E_n)^2+\Gamma_n^2} .
\end{equation} 
Here, $\Gamma_n$ denotes the fermion damping rate \cite{Ghosh:2024hbf}. As elaborated in Sec.~\ref{sec:damping-rate}, a detailed knowledge of the damping rate as a function of the Landau-level index $n$ and longitudinal momentum $k_z$ is crucial for accurately determining the transport properties of a magnetized plasma.

We note that the expression for the correlator in Eq.~(\ref{sigma-tensor-spectral}) represents the leading-order result, where vertex corrections and ladder-diagram resummations of the self-energy are neglected. It can be argued that both effects yield only subleading contributions to the conductivity in strongly magnetized plasmas \cite{Hattori:2016cnt}. Qualitatively, this likely arises from the restricted kinematics associated with the infrared-sensitive lowest Landau level. Additionally, the corresponding damping rate of states in the lowest Landau level is governed by quantum transitions to other Landau levels, which reduces its sensitivity to infrared dynamics (see Sec.~\ref{sec:damping-rate} and Ref.~\cite{Ghosh:2024hbf} for details). This is in contrast to a plasma in the absence of a magnetic field, where such discrete Landau-level quantization does not occur.

The thermal conductivity tensor $\kappa_{ij}$ can be defined analogously to the electrical conductivity tensor $\sigma_{ij}$, but in terms of the correlation function of the energy currents rather than electric currents. For Dirac fermions, the energy current is given by $T^{i0}  = \frac{i}{4}\left(\bar{\psi} \gamma^i \partial_0 \psi - \partial_0 \bar{\psi} \gamma^i \psi+\bar{\psi} \gamma_0 \partial^i \psi - \partial^i \bar{\psi} \gamma_0 \psi\right)$ \cite{Goedecke:1974}, in contrast to the electric current $j^i = e \bar{\psi} \gamma^i \psi$ used in Eq.~(\ref{sigma-tensor}). When expressed in terms of the fermion spectral function $A_{\bm{k}}(k_0)$, the thermal conductivity tensor takes the form: 
\begin{eqnarray}
\kappa_{ij} &=& - \frac{1}{128\pi^2 T^2} \int \frac{d k_{0} d^3 \bm{k}}{\cosh^2\frac{k_{0}}{2T}}
\mbox{tr} \big[ k_{0}^2 \gamma^i A_{\bm{k}} (k_0) \gamma^j A_{\bm{k}} (k_0) +k^{i}k^{j} \gamma_0 A_{\bm{k}} (k_0) \gamma_0 A_{\bm{k}} (k_0)\nonumber\\
  &&+k_{0}k^{i} \gamma_0 A_{\bm{k}} (k_0) \gamma^j A_{\bm{k}} (k_0)
  +k_{0}k^{j} \gamma^i A_{\bm{k}} (k_0) \gamma_0 A_{\bm{k}} (k_0) \big].
\label{kappa-tensor-spectral}
\end{eqnarray}
It is worth emphasizing that, unlike its non-relativistic analog in Ref.~\cite{Ferrer:2002gf}, this expression includes an extra contribution, given by the second term in the integrand.

We assume that the fermion chemical potential is zero. In this case, it is easy to verify that the off-diagonal components of both the thermal and electrical conductivity tensors vanish. Furthermore, due to the rotational symmetry about the direction of the magnetic field, each tensor has only two independent components: the longitudinal components, $\kappa_\parallel$ and $\sigma_\parallel$, and the transverse components, $\kappa_\perp$ and $\sigma_\perp$, respectively. 
 
Substituting spectral density (\ref{spectral-density}) into Eqs.~(\ref{sigma-tensor-spectral}) and (\ref{kappa-tensor-spectral}), and performing the integration over $\bm{k}_\perp$, we obtain the final expressions for the transverse and longitudinal electrical conductivities,
\begin{subequations}
\begin{eqnarray}
\sigma_{\perp} &=& \frac{\alpha |eB|}{4T}
\sum_{n=0}^{\infty} \sum_{\{\lambda\}} 
\int \frac{d k_{0} d k_{z} }{\cosh^2\frac{k_{0}}{2T}}  
\rho\left(k_0,\lambda E_n\right)\rho\left(k_0,\lambda^\prime E_{n+1}\right)
\left(1-\lambda \lambda^\prime \frac{k_z^2+m^2}{E_n E_{n+1}}\right) ,
\label{sigma-11}
\\
\sigma_{\parallel} &=& \frac{\alpha |eB|}{8T}  \sum_{n=0}^{\infty} \sum_{\{\lambda\}}
\beta_n \int \frac{d k_{0} d k_{z} }{\cosh^2\frac{k_{0}}{2T}}  
\rho\left(k_0,\lambda E_n\right)\rho\left(k_0,\lambda^\prime E_n\right)
\left(1-\lambda \lambda^\prime +\lambda \lambda^\prime \frac{2k_z^2}{E_n^2}\right) ,
\label{sigma-33}
\end{eqnarray}
\end{subequations}
as well as for the transverse and longitudinal thermal conductivities,
\begin{subequations}
\begin{eqnarray}
\kappa_{\perp} &=&\frac{(eB)^2}{256\pi T^2}
\sum_{n=0}^{\infty}  \sum_{\lambda=\pm1} \int \frac{d k_{0} d k_{z} }{\cosh^2\frac{k_{0}}{2T}} 
\rho\left(k_0,\lambda E_n\right)^2 \left(4n+\delta_{n,0} +  \frac{8\lambda n k_0}{E_{n}} \right) 
\nonumber\\
&+& 
\frac{|eB|}{256\pi T^2}
\sum_{n=0}^{\infty} \sum_{\{\lambda\}} 
\int \frac{d k_{0} d k_{z} }{\cosh^2\frac{k_{0}}{2T}} 
\rho\left(k_0,\lambda E_n\right)\rho\left(k_0,\lambda^\prime E_{n+1}\right)
\Bigg[ 4k_0^2
\left(1-\lambda \lambda^\prime \frac{k_z^2+m^2}{E_n E_{n+1}}\right)
 \nonumber\\
&+&n|eB|\left( 1 + \lambda \lambda^{\prime} \frac{E_{n+1} }{E_{n} } + \frac{4\lambda k_0}{E_{n}}  \right) 
+(n+1)|eB| \left( 1 + \lambda \lambda^{\prime} \frac{E_{n} }{E_{n+1} } + \frac{4\lambda^\prime k_0}{E_{n+1}}   \right)    
\Bigg] ,
\label{kappa-11}
\\
\kappa_{\parallel} &=& \frac{|eB|}{128\pi T^2} \sum_{n=0}^{\infty} \sum_{\{\lambda\}} 
\beta_n  \int \frac{d k_{0} d k_{z} }{\cosh^2\frac{k_{0}}{2T}} 
\rho\left(k_0,\lambda E_n\right)\rho\left(k_0,\lambda^\prime E_n\right)\nonumber\\
&&\times
\Bigg[ 
k_z^2 \left(1+ \lambda \lambda^\prime  \right) \left(1+\frac{2 \lambda k_0}{E_n}\right) 
+ k_0^2 \left(1-\lambda \lambda^\prime +\lambda \lambda^\prime \frac{2k_z^2}{E_n^2}\right) 
\Bigg] ,
\label{kappa-33}
\end{eqnarray}
\end{subequations} 
respectively. Here, $\beta_n \equiv 2 - \delta_{n,0}$ accounts for the Landau-level spin degeneracy, the notation $\{\lambda\}$ denotes the sum over $\lambda=\pm 1$ and $\lambda^\prime=\pm 1$. 

It should be noted that the above analysis accounts only for the partial contributions of fermions to the thermal conductivity. Strictly speaking, however, one should also include a contribution of plasmons (i.e., medium-modified photons). The estimate of the latter can be obtained using the same arguments as in the absence of the magnetic field \cite{Hosoya:1983id,Shovkovy:2002kv}, 
\begin{equation}
\kappa_\gamma \simeq \frac{4\pi^2T^3}{45 \Gamma_\gamma} \sim \frac{T^4}{\alpha |eB|} ,
\label{kappa-photon}
\end{equation}
where we used an order of magnitude estimate for the photon damping rate, $\Gamma_\gamma\sim \alpha |eB|/T$. In the strong magnetic field limit, the photon contribution given in Eq.~(\ref{kappa-photon}) is suppressed and is therefore negligible compared to the fermionic contribution to the thermal conductivity.
 
 As evident from Eq.~(\ref{sigma-11}), the $n$th partial contribution to the transverse electrical conductivity is proportional to a product of Lorentzian functions peaked at two adjacent Landau levels (with $k_0 = \pm E_n$ and $k_0 = \pm E_{n+1}$). In the presence of a sufficiently strong magnetic field, and assuming weak interactions, these Lorentzians have little overlap because the damping rate is small compared to the Landau-level spacing. This regime naturally arises in a QED plasma when $\sqrt{|eB|} \lesssim T/\alpha$. Consequently, each partial contribution to the transverse electrical conductivity is suppressed. Physically, this reflects the fact that transverse charge transport is mediated by quantum transitions (or ``jumps") between adjacent Landau levels. In the limit of vanishing interaction strength, i.e., $\alpha \to 0$, these transitions are effectively forbidden. Then, fermions remain confined to their Landau orbits, and transverse transport becomes negligible.

In contrast, the behavior of the longitudinal electrical conductivity in Eq.~(\ref{sigma-33}) is very different. In this case, the partial contributions from each Landau level are dominated by products of Lorentzian functions centered at the same Landau-level energy (when $\lambda = \lambda^\prime$), leading to significantly larger contributions that scale as $1/\alpha$. In the weak-coupling limit, these contributions become large, indicating that particles within each Landau level (effectively behaving as distinct species) can move freely along the direction of the magnetic field with little scattering. This charge transport mechanism closely resembles that in the absence of a magnetic field. However, when $B \neq 0$, the longitudinal conductivity is further enhanced by two other effects: the increased density of states in each Landau level, which is proportional to $|eB|$, and the suppression of backscattering. Both effects arise from the quasi-one-dimensional nature of charged particles in a strong magnetic field, which is commonly referred to as dimensional reduction \cite{Miransky:2015ava}.
 
 Both transverse and longitudinal thermal conductivities are dominated by terms with products of Lorentzian functions centered at the same Landau-level energy (for $\lambda = \lambda^\prime$), closely resembling the structure found in the longitudinal charge conductivity. This indicates that, unlike charge transport, transverse heat transport is not suppressed relative to its longitudinal component. The physical interpretation of this behavior is connected with a different mechanism and will be discussed in more detail in Sec.~\ref{sec:transport-QED}.
 
\section{Fermion damping rate}
\label{sec:damping-rate}

To compute the electrical conductivities given by Eqs.~(\ref{sigma-11}) and (\ref{sigma-33}) and the thermal conductivities given by Eqs.~(\ref{kappa-11}) and (\ref{kappa-33}), it is essential to determine the damping rates $\Gamma_n(k_z)$ of charge carriers in a strongly magnetized plasma. These damping rates set the widths of the Lorentzian peaks in the fermion spectral function and thus play a crucial role in shaping the transport properties. Using the Landau-level representation, the relevant expressions for $\Gamma_n(k_z)$ were derived in Ref.~\cite{Ghosh:2024hbf} within a gauge theory framework at leading order in the coupling.
 
\begin{figure}[t]
\centering
  \subfigure[]{\includegraphics[width=0.23\textwidth]{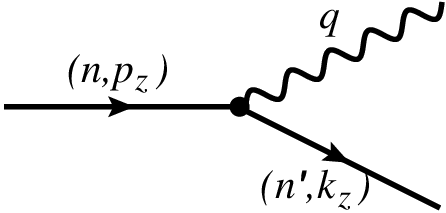}}
  \hspace{0.1\textwidth}
  \subfigure[]{\includegraphics[width=0.23\textwidth]{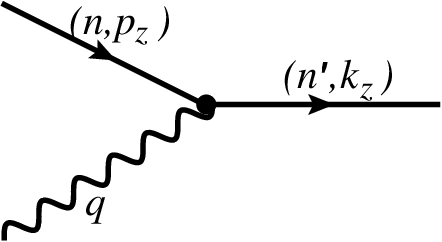}}
  \hspace{0.1\textwidth}
  \subfigure[]{\includegraphics[width=0.23\textwidth]{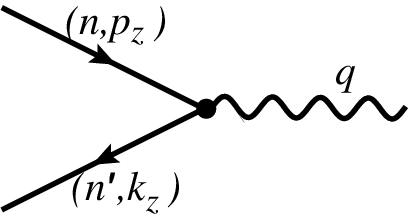}}
\caption{Leading order processes contributing to the fermion damping rates:
(a) $\psi_{n}\to \psi_{n^\prime}+\gamma$ with $n >n^{\prime}$, (b) $\psi_{n}+\gamma\to\psi_{n^{\prime}}$ with $n<n^{\prime}$, (c) $\psi_{n}+\bar{\psi}_{n^{\prime}}\to\gamma$, where $n$ and $n^{\prime}$ are the Landau-level indices.}
\label{Gamma-processes}
\end{figure}

Assuming a sufficiently strong magnetic field, specifically $|eB| \gg \alpha T^2$, the damping rate $\Gamma_n(k_z)$ of a quantum particle in a given Landau-level state is governed by the three types of processes represented by the Feynman diagrams in Fig.~\ref{Gamma-processes}: (i) downward transitions to Landau levels with lower indices $n^\prime$  ($\psi_{n}\to \psi_{n^\prime} +\gamma$ with $n>n^\prime$), (ii) upward transitions to Landau levels with higher indices $n^\prime$ ($\psi_{n}+\gamma\to \psi_{n^\prime} $ with $n<n^\prime$), and (iii) 
annihilation with negative-energy states ($\psi_{n}+\bar{\psi}_{n^\prime}\to \gamma$ for any $n$ and $n^\prime$). 

Note that the damping rate of a state in the $n$th Landau level is determined by transitions to other levels ($n^\prime > n$ or $n^\prime < n$), as well as by annihilation with antiparticle states in any Landau level. We argue that this is important for ensuring that self-energy resummations contribute only subleading corrections to the conductivity of a plasma in a strong magnetic field. In particular, transitions to other levels regulate the infrared dynamics, which would otherwise exhibit undue sensitivity in a thermal plasma at vanishing magnetic field \cite{Gagnon:2006hi}.

The resulting spin-averaged damping rate in the $n$th Landau level is given by the following exact expression obtained at the leading order in coupling \cite{Ghosh:2024hbf}:
 \begin{equation}
  \Gamma_{n}(k_z) = \frac{\alpha |eB|}{2 \beta_n  E_{n}} 
  \sum_{n^{\prime}=0}^\infty   \sum_{\{s\}}  
  \int   \frac{d\xi  \left[ 1-n_F(E_{f,s^\prime})+n_B(E_{\gamma,s^\prime}) \right]  {\cal M}_{n,n^{\prime}} (\xi) }{s_1 s_2 \sqrt{(\xi-\xi_{n,n^{\prime}}^{-})(\xi-\xi_{n,n^{\prime}}^{+} )} } , 
  \label{Gamma_n}
 \end{equation}
where $\xi_{n,n^{\prime}}^{\pm}=\frac{1}{2}\left[\sqrt{2n^{\prime}+(m\ell)^2}\pm \sqrt{2n+(m\ell)^2}\right]^2$ are the dimensionless transverse-momentum threshold parameters, $n_F(E)$ and $n_B(E)$ are the Fermi-Dirac and Bose-Einstein distribution functions, respectively. The shorthand notation $\{s\}$ denotes summation over $s^\prime=\pm 1$, $s_1=\pm 1$, and $s_2=\pm 1$. The function ${\cal M}_{n,n^{\prime}} (\xi)$ is determined by the squared amplitude of the leading-order processes shown in Fig.~\ref{Gamma-processes}. It is explicitly given by
\begin{eqnarray}
{\cal M}_{n,n^{\prime}}(\xi)  &=&-  \left(n+n^{\prime}+ (m\ell)^2\right)\left[\mathcal{I}_{0}^{n,n^{\prime}}(\xi)+\mathcal{I}_{0}^{n-1,n^{\prime}-1}(\xi) \right]  \nonumber\\
&&
  +(n+n^{\prime}) \left[\mathcal{I}_{0}^{n,n^{\prime}-1}(\xi)+\mathcal{I}_{0}^{n-1,n^{\prime}}(\xi) \right] ,
  \label{Mnnp}
 \end{eqnarray} 
where the form factor function $\mathcal{I}_{0}^{n,n^{\prime}}(\xi)$ is the same as in Ref.~\cite{Wang:2021ebh}: 
 \begin{equation}
  \mathcal{I}_{0}^{n,n^{\prime}}(\xi) =
  \frac{(n^\prime)!}{n!} e^{-\xi}  \xi^{n-n^\prime} \left(L_{n^\prime}^{n-n^\prime}\left(\xi\right)\right)^2   \label{I0}  .
 \end{equation} 
 In Eq.~(\ref{Gamma_n}), the integration range for $\xi$ depends on the signs of $s_1$ and $s_2$: (i) for $s_1>0$ and any sign of $s_2$, the range is $0<\xi<\xi_{n,n^{\prime}}^{-}$ and (ii) for $s_1<0$ and $s_2>0$, the range is $\xi_{n,n^{\prime}}^{+}<\xi< \infty$. The combination $s_1 < 0$ and $s_2 < 0$ does not contribute to the damping rates of particles with positive energy and can be ignored.
  
The energies of the other fermion ($E_{f,s^\prime}$) and the photon ($E_{\gamma,s^\prime}$) in Eq.~(\ref{Gamma_n}) are 
	\begin{subequations}
 \begin{eqnarray}
E_{f,s^\prime}  &=& 
  \frac{E_{n}}{2}\left(1+\frac{2n^{\prime}+(m\ell)^2-2\xi}{2n +(m\ell)^2}  \right)
 + s^\prime k_z \frac{\sqrt{(\xi-\xi_{n,n^{\prime}}^{-})(\xi-\xi_{n,n^{\prime}}^{+} )}}{ 2n+(m\ell)^2} ,
  \label{Ef-mass} \\
E_{\gamma,s^\prime} &=& 
  \frac{E_{n}}{2}\left(1-\frac{2n^{\prime}+(m\ell)^2-2\xi}{2n +(m\ell)^2}  \right)
- s^\prime k_z \frac{\sqrt{(\xi-\xi_{n,n^{\prime}}^{-})(\xi-\xi_{n,n^{\prime}}^{+} )}}{ 2n+(m\ell)^2} ,
  \label{Egamma-mass}
 \end{eqnarray}
 	\end{subequations}
 respectively. These expressions result from solving the energy conservation conditions for the relevant one-to-two and two-to-one processes.
 
 In the chiral limit ($m = 0$), however, the corresponding energies for the lowest Landau level ($n = 0$) are
 	\begin{subequations}
 \begin{eqnarray}
 E_{f}  &=& - \frac{(\xi -n^{\prime})^2 +2n^{\prime} k_z^2\ell^2 }{2\ell^2 |k_z|(\xi-n^{\prime})} , 
  \label{Ef-chiral} \\
E_{\gamma}  &=&  \frac{|eB|(\xi -n^{\prime})^2 +2\xi k_z^2}{2 |k_z| (\xi -n^{\prime})} ,
  \label{Egamma-chiral} 
 \end{eqnarray}
 	\end{subequations}
respectively. Unlike Eqs.~(\ref{Ef-mass}) and (\ref{Egamma-mass}), which allow for two solutions labeled by $s' = \pm 1$, only a single solution exists for $n = 0$ in the chiral limit.
 
\section{Heat and charge transport in QED}
\label{sec:transport-QED}

In the case of the electron-positron plasma, it is straightforward to compute the detailed numerical dependence of $\Gamma_{n}(k_z)$ on both $n$ and $k_z$, using the definition of the damping rate in Eq.~(\ref{Gamma_n}). With this data, we can readily calculate the transverse and longitudinal transport coefficients, as defined in Eqs.~(\ref{sigma-11}) through (\ref{kappa-33}). The corresponding results for the electrical conductivity were reported in Refs.~\cite{Ghosh:2024fkg,Ghosh:2024owm}. Here, using tabulated numerical data for damping rates, we further extend those results by evaluating the thermal conductivity.\footnote{In this work, we employ a larger set of damping-rate data and include additional Landau levels but reduce slightly the resolution in the longitudinal momentum $k_z$, compared with previous articles.}

\begin{figure}[t]
\centering
\includegraphics[width=0.48\columnwidth]{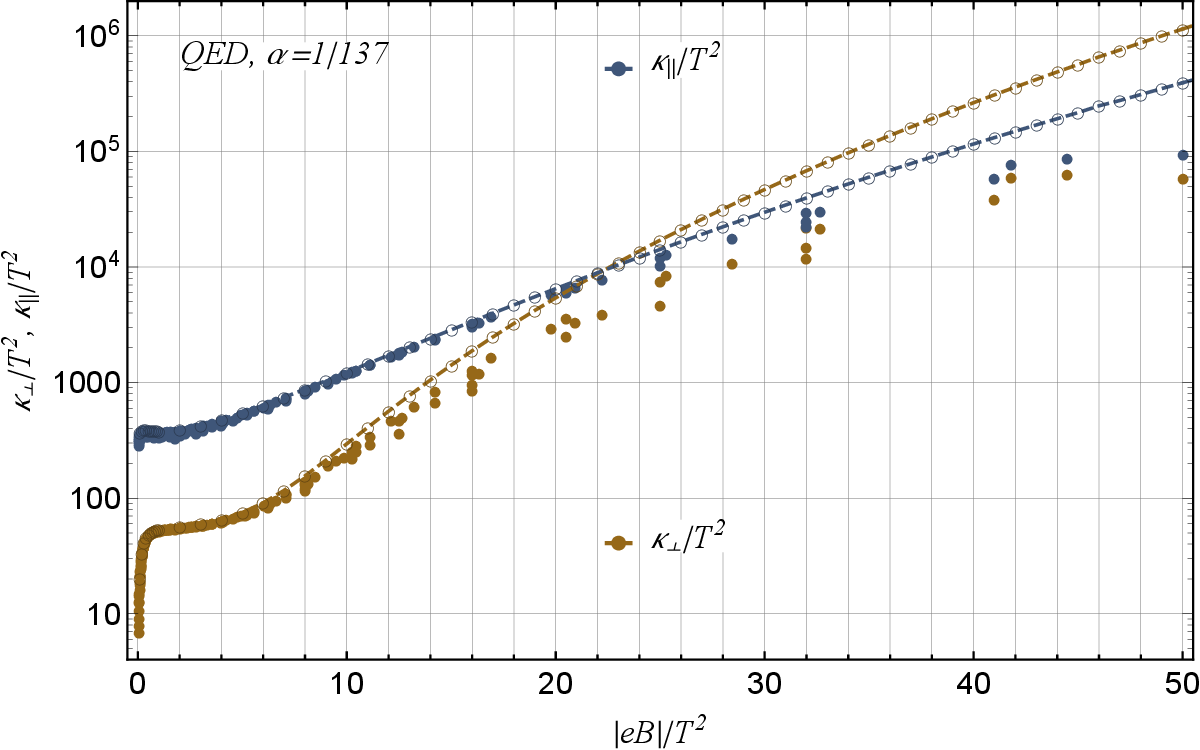} 
\hspace{0.02\columnwidth}
\includegraphics[width=0.48\columnwidth]{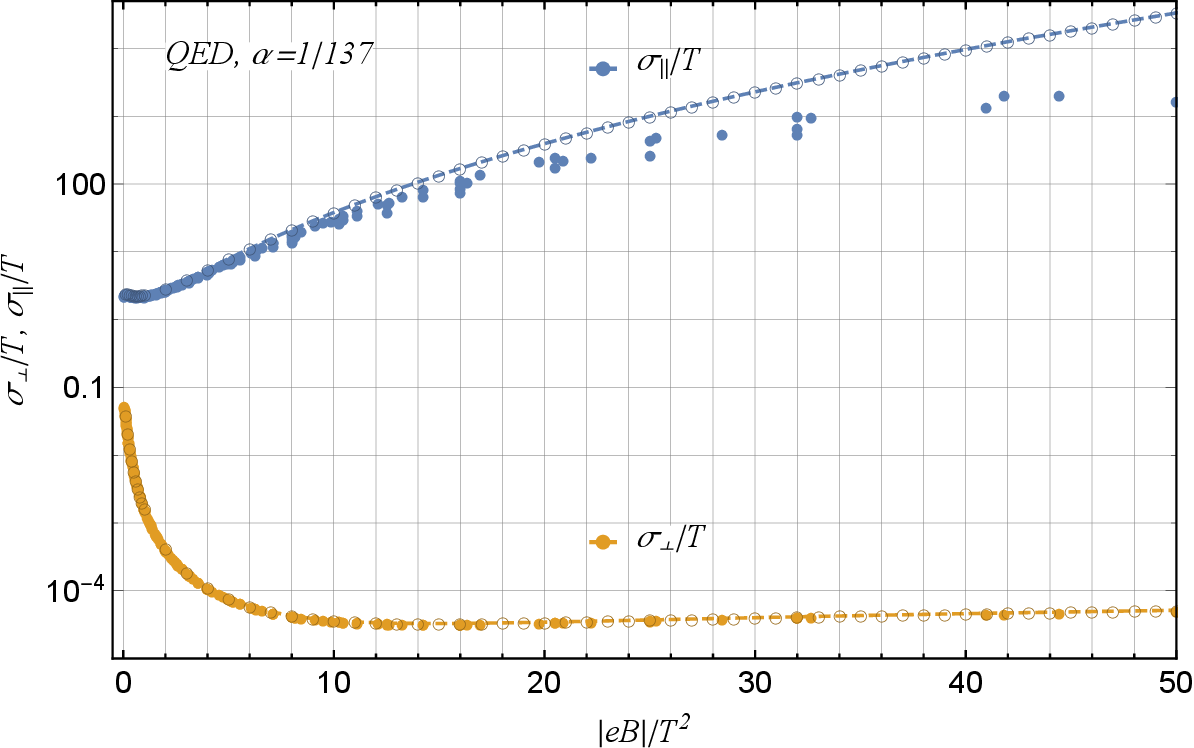} 
\caption{The thermal (left) and electrical (right) conductivities as functions of the dimensionless ratio $|eB|/T^2$. Empty circles and interpolating dashed lines represent the results in the chiral limit.}
\label{fig:res_QED}
\end{figure}

The results for thermal conductivity are summarized in the left panel of Fig.~\ref{fig:res_QED}, where we show the dimensionless ratios $\tilde{\kappa}_{\perp}\equiv \kappa_{\perp}/T^2$ (brown) and $\tilde{\kappa}_{\parallel}\equiv \kappa_{\parallel}/T^2$ (dark blue) as functions of $|eB|/T^2$. They are compared with the results for the electrical conductivity in the right panel of Fig.~\ref{fig:res_QED}, where we show analogous dimensionless ratios, i.e., $\tilde{\sigma}_{\perp}\equiv \sigma_{\perp}/T$ (orange) and $\tilde{\sigma}_{\parallel}\equiv \sigma_{\parallel}/T$ (blue) as functions of $|eB|/T^2$. Filled markers correspond to results obtained with a finite fermion mass, while open markers represent the chiral limit ($m = 0$). 

In plotting Fig.~\ref{fig:res_QED}, we included a large set of data for a wide range of high temperatures, $T\gg m$, and strong magnetic fields, $|eB|\gg m^2$, where $m\approx 0.511~\mbox{MeV}$ is the electron mass. Specifically, the temperatures span from $20m$ to $85m$, while the magnetic field ranges from $(20m)^2$ to $(226m)^2$. (Numerical data are provided in the Supplemental Material \cite{DataCond:2025}.) In this relativistic regime, it is reasonable to expect that the effects of a nonzero fermion mass are small. This motivates the conjecture that the dimensionless transport coefficients (i.e., conductivities normalized by appropriate powers of temperature) are approximately universal functions of the dimensionless ratio $|eB|/T^2$. This expectation is largely supported by the data shown in Fig.~\ref{fig:res_QED}. The close agreement between the results with and without a finite fermion mass indicates that mass effects are indeed relatively small. It should be noted, however, that deviations from the expected scaling behavior are negligible for $\sigma_{\perp}$ but more noticeable in other transport coefficients, where sensitivity to a nonzero mass is significantly enhanced. Furthermore, we find that a finite resolution of the tabulated numerical data for $\Gamma_{n}(k_z)$, especially at small values of $k_z$, introduces a systematic error that tends to underestimate the longitudinal conductivity. This effect becomes more pronounced as $|eB|/T^2$ increases. One can argue, however, that the correct value should be bounded from above by the results in the chiral limit.

It is instructive to examine more closely the underlying physics of longitudinal and transverse transport in the strong-field limit, where $|eB|/T^2 \gg 1$. Naively, one may expect this regime to be dominated by particles in the lowest Landau level. This expectation largely holds true for the longitudinal conductivities $\kappa_{\parallel}$ and $\sigma_{\parallel}$, as well as $\kappa_{\perp}$, where particles in the $n = 0$ Landau level contribute the dominant share to transport. However, it is important to recall that the damping rates of states in the lowest Landau level are not determined by processes within that level. As seen from the definition in Eq.~(\ref{Gamma_n}), $\Gamma_{0}(k_z)$ receives contributions from transitions to higher Landau levels, as well as from annihilation with antiparticles across all Landau levels (including the lowest).

In contrast, transverse charge transport in the strong-field limit is more subtle. Since the partial contributions to transverse electrical conductivity $\sigma_{\perp}$ arise from quantum transitions between adjacent Landau levels, the first excited level ($n = 1$) plays a role that is at least as important as that of the lowest level. In fact, the transverse conductivity formally vanish if contributions from higher Landau levels ($n > 0$) are excluded from the dynamics. 

The behavior of transverse thermal conductivity $\kappa_{\perp}$ formally resembles that of longitudinal transport ($\kappa_{\parallel}$ and $\sigma_{\parallel}$). Indeed, its dominant contributions come from terms with products of Lorentzian functions centered at the same Landau-level energy. This is evident from the structure of Eq.~(\ref{kappa-11}). Such behavior is rather unusual and stands in sharp contrast to transverse charge transport. The interpretation can be traced to the underlying nature of heat transport, which is associated with the dissipative components of collective modes. Since this transport via collective modes does not involve charge transfer, it is not constrained by the magnetic field.

\section{Heat and charge transport in QGP}
\label{sec:transport-QGP}

In contrast to the QED plasma, the quark-gluon plasma is composed of strongly interacting quarks and gluons, rendering perturbative approaches generally unreliable in the temperature regimes accessible in current heavy-ion collision experiments. As a result, leading-order calculations of the damping rate due to one-to-two and two-to-one processes cannot be expected to yield quantitatively accurate predictions. Nevertheless, such perturbative estimates for the damping rate and, by extension, for transport properties, remain valuable as theoretical benchmarks. They provide insights into the qualitative behavior of transport coefficients and serve as a starting point for understanding the parametric dependence of these quantities on key variables such as temperature and magnetic field strength. Furthermore, this perturbative approach becomes justified in regimes of extremely high temperature \cite{Busza:2018rrf}, where the QCD coupling becomes weak due to asymptotic freedom \cite{Gross:1973id,Politzer:1973fx}.

To extend the QED calculation of thermal and electrical conductivities, as given by Eqs.~(\ref{sigma-11}) through (\ref{kappa-33}), to QGP, several important modifications must be made. First, instead of a single electron, one must account for all quark flavors and colors contributing to transport. In the expression for electrical conductivity, the correct quark charges must be used: $e_f = q_f e$, where $q_u = 2/3$ and $q_d = -1/3$ for up and down quarks, respectively. Additionally, the corresponding expressions in Eqs.~(\ref{sigma-11}) through (\ref{kappa-33}) should be multiplied by an extra factor of $N_c=3$ to account for the quark colors. (Note that the magnetic length for the up and down quarks, appearing in their spectral densities, are also different.) As in the QED case, only the quark contributions are considered here. In general, however, medium-modified gluons (as well as photons) are also expected to contribute to thermal conductivity. Their contribution is estimated to be of order $T^4/(\alpha_s |eB|)$, governed by the gluon damping rate $\Gamma_g \sim \alpha_s |eB|/T$. In the strong magnetic field limit, this gluonic contribution is expected to be negligible compared to the thermal conductivity from quarks.

In addition to these explicit modifications of the conductivity expressions in Eqs.~(\ref{sigma-11}) through  (\ref{kappa-33}), an important implicit change arises from the quark damping rate. In QCD, the damping rates $\Gamma_n^f(k_z)$ are also governed by one-to-two and two-to-one processes. Unlike in QED, these processes involve gluons rather than photons: $\psi_n \to \psi_{n'} + g$, $\psi_n + g \to \psi_{n'}$, and $\psi_n + \bar{\psi}_{n'} \to g$, where $g$ denotes a gluon. The corresponding definition for the rate is similar to that in Eq.~(\ref{Gamma_n}), but the coupling constant $\alpha$ should be replaced with $\alpha_s C_F$, where $\alpha_s = g_s^2/(4\pi)$ is the strong coupling constant and $C_F = (N_c^2 - 1)/(2N_c) = 4/3$, assuming $N_c = 3$ \cite{Ghosh:2024hbf}.

\begin{figure}[t]
  \centering
  \includegraphics[width=0.48\columnwidth]{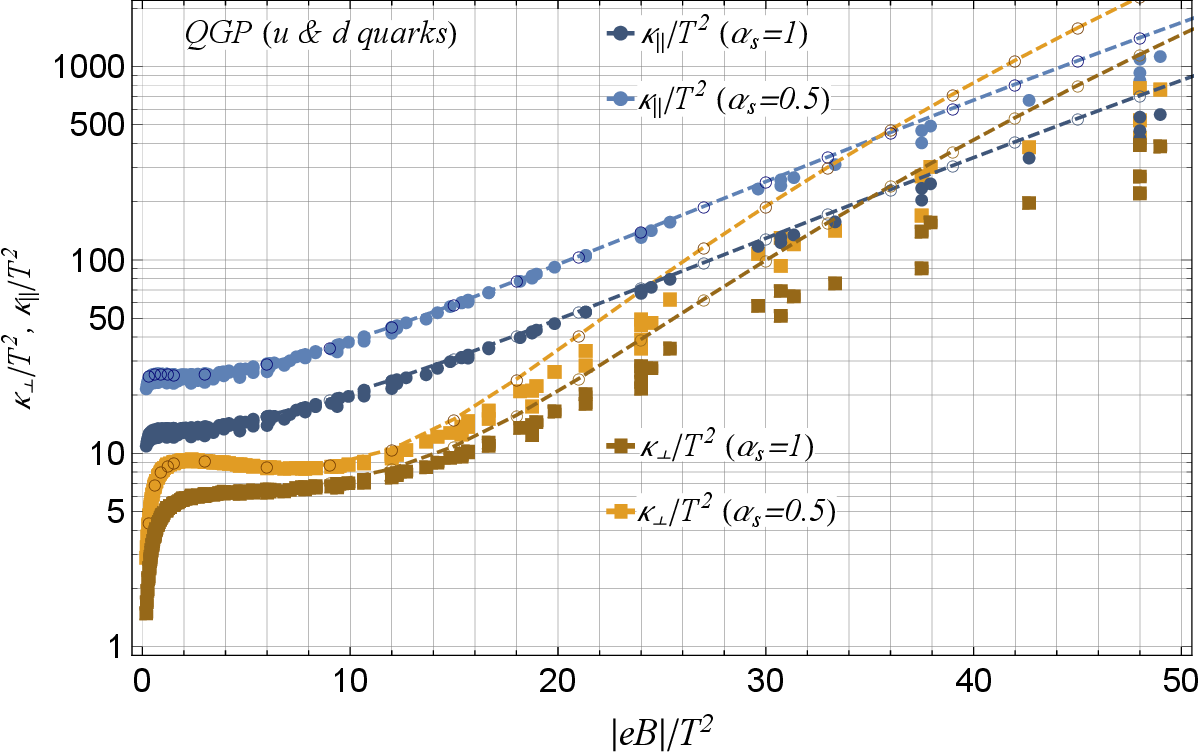} 
  \hspace{0.02\columnwidth}
  \includegraphics[width=0.48\columnwidth]{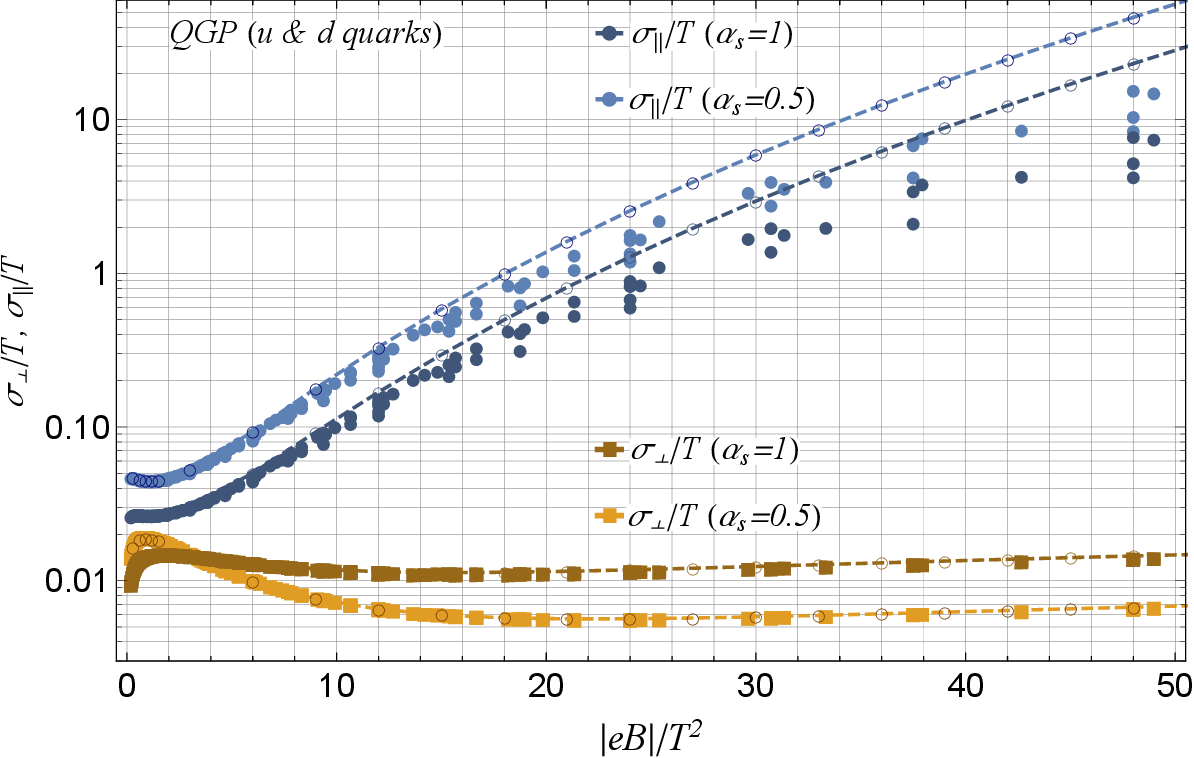} 
  \caption{The thermal (left) and electrical (right) conductivities of two-flavor QGP as functions of the dimensionless ratio $|eB|/T^2$. Empty circles and interpolating dashed lines represent the results in the chiral limit.}
  \label{fig:res_QCD}
 \end{figure}

Given that the typical temperatures of the QGP produced in ultrarelativistic heavy-ion collisions are on the order of several hundred megaelectronvolts, the strong coupling constant $\alpha_s$ is generally of order unity. To obtain numerical estimates for the transport coefficients, we consider two representative values of the coupling constant: $\alpha_s = 0.5$ and $\alpha_s = 1$. The corresponding results for the thermal and electrical conductivities are shown in Fig.~\ref{fig:res_QCD}. The plot includes many data points for a broad range of high temperatures, $20m \lesssim T \lesssim 80m$, and magnetic field strengths, $(15m)^2 \lesssim |eB| \lesssim (225m)^2$, where $m \approx 5~\mbox{MeV}$ denotes the quark mass. (Numerical data are provided in the Supplemental Material \cite{DataCond:2025}.) Within this parameter range, the effects of a finite quark mass are expected to be small. Accordingly, as in the QED case shown in Fig.~\ref{fig:res_QED}, the dimensionless transport coefficients approximately follow universal scaling behaviors as functions of the ratio $|eB|/T^2$.

For comparison, the results in the chiral limit are indicated by open markers and interpolated with dashed lines in Fig.~\ref{fig:res_QCD}. As evident from the figure, the transverse electrical conductivity closely matches its chiral-limit counterpart, showing minimal sensitivity to the quark mass. In contrast, the longitudinal transport coefficients, as well as   
$\kappa_{\perp}$, exhibit significantly larger deviations from the chiral limit. This sensitivity is primarily an artifact of the lowest Landau level quarks, whose damping rates depend moderately strongly on the quark mass. (As in QED, a finite resolution of the tabulated $\Gamma_n^f(k_z)$, especially at small values of $k_z$, leads to a systematic underestimate of $\sigma_{\parallel}$ and both components of the thermal conductivity. Nevertheless, the correct values should remain well below their counterparts in the chiral limit.)

As previously emphasized, the leading-order approximation is not reliable for quantitatively estimating the transport coefficients in a strongly interacting QGP. It is therefore not surprising that the results for the electrical conductivities presented in Fig.~\ref{fig:res_QCD} differ substantially from those obtained in lattice QCD simulations \cite{Astrakhantsev:2019zkr,Almirante:2024lqn}. While the transverse electrical conductivity shows reasonable agreement with lattice results, the longitudinal conductivity is significantly larger (for details, see Figs. 6 and 7 in Ref.~\cite{Ghosh:2024owm}). In the limit of very strong magnetic fields, it differs by several orders of magnitude. This discrepancy likely reflects the importance of higher-order contributions and, more generally, nonperturbative effects that are not captured within the leading-order framework.

\section{Deviations from the Wiedemann--Franz law}

The Lorenz number is defined as the ratio of thermal conductivity to the product of electrical conductivity and temperature, $L=\kappa/(T\sigma )$. In metals, where both heat and charge are transported by the same quasiparticles, this ratio approaches a universal value, $L_0=\pi^2k_B^2/(3e^2)$, which encapsulates the essence of the Wiedemann--Franz law \cite{Ashcroft-Mermin:1976}. However, there is no fundamental reason to expect this relation to hold in a relativistic plasma. Indeed, strong violations of the Wiedemann--Franz law have been observed in systems such as graphene at the charge neutrality point \cite{Fong:2013PRX041008,Crossno-Fong:2016} and in the Weyl semimetal tungsten diphosphide \cite{Gooth-Felser:2018}. Even in conventional metals, the Wiedemann--Franz law breaks down in the presence of inelastic scattering of quasiparticles \cite{Ashcroft-Mermin:1976}. In a relativistic plasma subjected to a strong magnetic field, the fermion damping rate is dominated by inelastic one-to-two and two-to-one processes, which fundamentally violate the assumptions underlying the Wiedemann--Franz law. Consequently, there is no reason to expect the Lorenz number to approach any universal value in such a system. Nevertheless, it remains instructive to ask: What is the Lorenz number in a strongly magnetized plasma?

\begin{figure}[t]
  \centering
  \includegraphics[width=0.48\columnwidth]{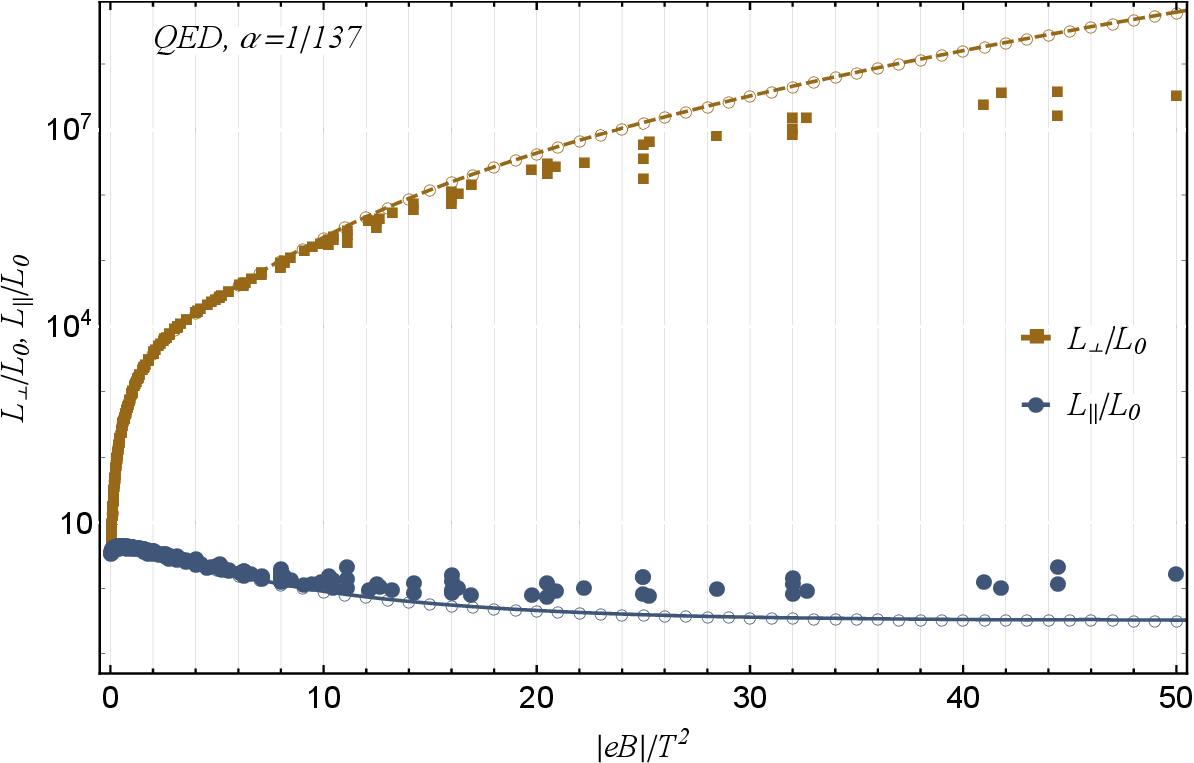} 
  \hspace{0.02\columnwidth}
 \includegraphics[width=0.48\columnwidth]{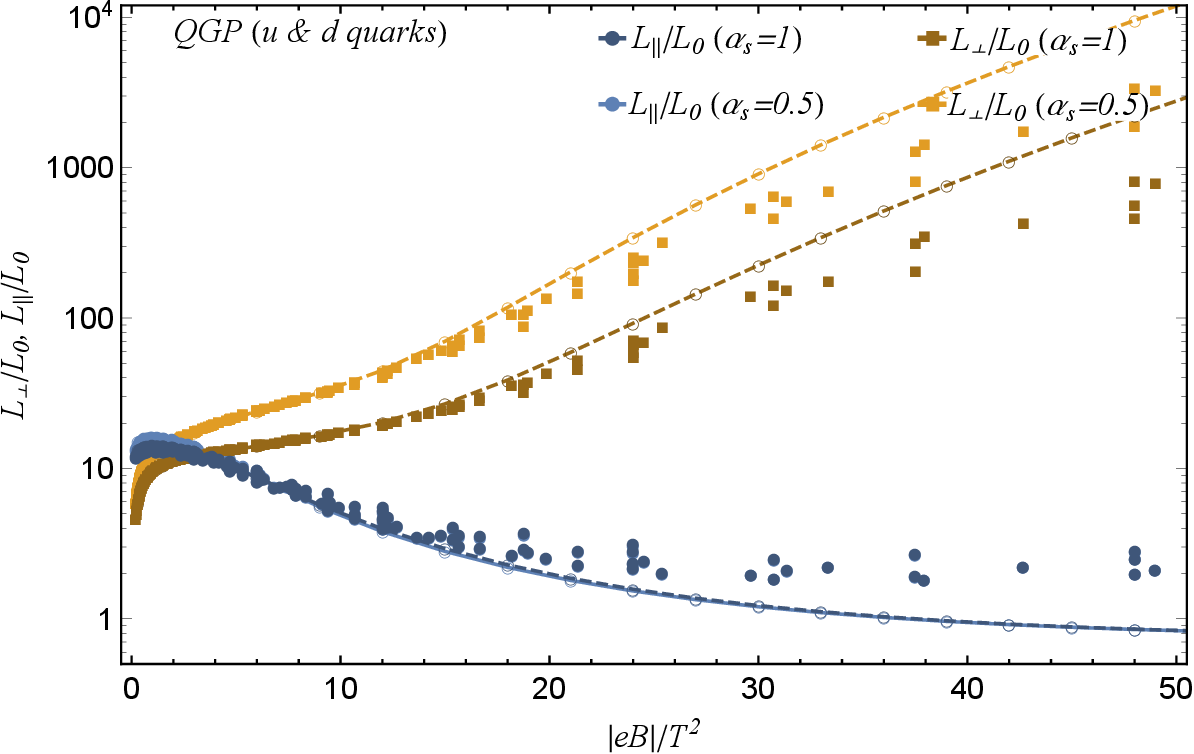} 
  \caption{The transverse and longitudinal Lorenz numbers as functions of the dimensionless parameter $|eB|/T^2$ for QED plasma (left) and QGP (right). Results in the chiral limit are shown by empty circles and interpolating dashed lines. In the case of QGP (right), the results for two different choices of the coupling constant, $\alpha_s=0.5$ and $\alpha_s=1$,  are shown.}
  \label{fig:ratioL_QED}
 \end{figure}

In the presence of the background magnetic field, two distinct Lorenz numbers for the longitudinal and transverse conductivities can be defined, i.e., 
\begin{equation}
L_{\perp,\parallel}= \frac{\kappa_{\perp,\parallel}}{ T \sigma_{\perp,\parallel}} .
\end{equation}
Figure~\ref{fig:ratioL_QED} shows the numerical results for the Lorenz numbers, obtained using the same data points as in Figs.~\ref{fig:res_QED} and \ref{fig:res_QCD}. As anticipated, both the longitudinal and transverse Lorenz numbers exhibit a nontrivial dependence on the model parameters, underscoring the non-universality of the relation between heat and charge transport in a strongly magnetized plasma.

In the chiral limit, the longitudinal Lorenz numbers approach fixed values in the strong magnetic field regime, where $|eB|/T^2\gg 1$. For chiral QED, the limiting value is given approximately by $L_{\parallel} \approx L_0/3$. This behavior suggests a potentially universal feature in the weakly coupled regime in strongly magnetized plasmas at zero chemical potential ($\mu=0$). In contrast, the transverse Lorenz number $L_{\perp}$ exhibits markedly different behavior, increasing with the magnetic field. This trend naturally reflects the underlying conductivities: the transverse thermal conductivity grows with the field, while the transverse electrical conductivity decreases. When $m \neq 0$, both longitudinal and transverse Lorenz numbers exhibit substantial variability, even at large $|eB|/T^2 \gg 1$. The absence of clear convergence may be attributed to limitations of the dataset, which includes multiple temperatures, which are not sufficiently high compared to the fermion mass, as well as systematic errors introduced by the finite discretization of numerical damping rates as functions of the longitudinal momentum, especially at small values of $k_z$.

A qualitatively similar behavior emerges also in QCD. In the chiral limit, the longitudinal Lorenz number approaches $L_{\parallel}\approx 0.76L_0$ in the strong-field regime, $|eB|/T^2\gg 1$. Despite strong coupling, this appears to be surprisingly consistent with the QED result after accounting for different quark charges. Indeed, by separating their individual contributions in the chiral limit, we verified that their flavor-specific Lorenz longitudinal numbers approach similar relations: $L_{f,\parallel} \approx L_0/(3q_f^2)$, where $f=u,d$ denotes the quark flavor. As for the transverse Lorenz number, it increases with the magnetic field as in QED. 

In summary, our results in both QED and QCD suggest that, in the strong magnetic field regime $|eB|/T^2\gg 1$, the single-flavor longitudinal Lorenz number approaches a universal limit: $L_{f,\parallel} \approx L_0/(3q_f^2)$. Although transport in this regime is dominated by inelastic one-to-two and two-to-one processes, it is reasonable to expect that these processes affect longitudinal heat and charge transport in a similar manner. As a result, the corresponding Lorenz number approaches a fixed value. In contrast, the transverse Lorenz number grows as a function of $|eB|/T^2$, reflecting the fundamentally different nature of heat and charge transport across the magnetic field.

\section{Summary}

In this review, we examined the transport of heat and electric charge in relativistic plasmas exposed to strong magnetic fields. Such plasmas play important roles in the context of the early universe, neutron stars, and relativistic heavy-ion collisions. While our primary focus was on transport phenomena in QED plasmas, we also extended the analysis to the quark-gluon plasma, providing insights into the qualitative similarities and key differences between the two cases.

Using tools from quantum field theory, specifically Kubo's formalism in conjunction with fermion spectral functions derived from first principles in the Landau-level representation, we calculated the anisotropic thermal and electrical conductivities of relativistic plasmas in the presence of strong magnetic fields. A crucial input in this framework is the fermion damping rate, which captures how particles lose coherence due to inelastic interactions. These are governed by one-to-two and two-to-one processes: $\psi_{n}\to \psi_{n^\prime}+\gamma$, $\psi_{n}+\gamma \to \psi_{n^\prime}$,  and $\psi_{n}+\bar{\psi}_{n^\prime}\to \gamma$ \cite{Ghosh:2024hbf}. The rates in the Landau-level representation are nontrivial functions of the Landau-level index and the longitudinal momentum.

The study shows that magnetic fields strongly affect transport. Charge flow along the magnetic field is enhanced, while flow across the field is heavily suppressed due to Landau-level trapping, which restricts particle motion in directions perpendicular to the field. We also argued that these transport properties generally depend on the strength of the magnetic field and the temperature through a simple dimensionless ratio, $|eB|/T^2\gg 1$, especially in the ultrarelativistic regime where particle masses are small compared to the energy scales set by the temperature and magnetic field. While this scaling behavior is exact in the chiral limit, deviations arise in the presence of a nonzero fermion mass. In general, the scaling is more robust for transverse electrical conductivity and more sensitive to fermion mass in the longitudinal channel.

Longitudinal heat transport closely resembles charge transport. In contrast, transverse heat transport behaves quite differently. It increases rapidly as a function of $|eB|/T^2$, resembling longitudinal transport. This behavior reflects the unusual nature of thermal conductivity, which is governed by the dissipative components of collective sound-like modes and is not suppressed in the direction perpendicular to the magnetic field.

Extending the analysis to QGP, we found qualitatively similar features in the behavior of conductivities, though the absolute values differ significantly from lattice QCD results \cite{Astrakhantsev:2019zkr,Almirante:2024lqn}, especially for the longitudinal conductivity. This discrepancy underscores the importance of higher-order and nonperturbative effects that are not captured within the leading-order formalism employed in this study. Future work should aim to incorporate subleading contributions, such as two-to-two scattering processes, as well as nonperturbative effects, to improve the quality of theoretical transport studies in QGP. 

Finally, we examined deviations from the Wiedemann--Franz law in strongly magnetized plasmas. Our results reveal that the Lorenz numbers exhibit a nontrivial dependence on model parameters and the magnetic field strength, signaling a clear breakdown of the conventional relationship between thermal and electrical conductivities in these systems. In the chiral limit, however, we find that the flavor-specific longitudinal Lorenz numbers approach universal limiting values, $L_{f,\parallel} \approx L_0/(3q_f^2)$, in the regime of sufficiently strong magnetic fields. Remarkably, these limiting values remain approximately unchanged even at strong coupling for each species of charged fermions. This suggests that the inelastic one-to-two and two-to-one processes, which dominate the damping rate when $|eB|/T^2\gg 1$, govern both longitudinal heat and charge transport in a similar manner, thereby leading to fixed values for the Lorenz numbers. By contrast, the transverse Lorenz numbers in both QED and QCD are increasing functions of $|eB|/T^2$, driven primarily by the growth of the transverse thermal conductivity. While this qualitative behavior is understood, the precise underlying mechanism remains unclear and warrants further investigation in future studies.

\backmatter

\bmhead{Acknowledgements}
I.~A.~S. is grateful to Matthias Kaminski, Pavel Kovtun and Guy D. Moore for insightful discussions on the transport properties of relativistic plasmas. This work was supported in part by the U.S. National Science Foundation under Grant Nos.~PHY-2209470 and PHY-2514933.

\section*{Statements and Declarations}

\begin{itemize}
\item {\bf Availability of Data and Material:} The numerical data for the thermal and electrical conductivity of hot
  relativistic plasmas in a strong magnetic field are available as the Supplemental Material in Ref.~\cite{DataCond:2025}.
\item {\bf Funding:} This research was funded in part by the U.S. National Science Foundation under Grant Nos.~PHY-2209470  and PHY-2514933.
\item {\bf Conflict of interest/Competing interests:} The authors declare no conflicts of interest or competing interests relevant to this work.
\item {\bf Author contribution:} I.S. and R.G. jointly developed the theoretical framework, carried out the analytical derivations, and performed the numerical calculations. I.S. prepared the initial draft of the manuscript. Both authors contributed to discussions of the results and reviewed and approved the final version of the paper.
\end{itemize}





\end{document}